\documentclass[twoside,twocolumn,9pt]{article}
\usepackage{extsizes}
\usepackage[super,sort&compress,comma]{natbib}
\usepackage[version=3]{mhchem}
\usepackage[left=1.5cm, right=1.5cm, top=1.785cm, bottom=2.0cm]{geometry}
\usepackage{balance}
\usepackage{mathptmx}
\usepackage{sectsty}
\usepackage{graphicx}
\usepackage{lastpage} 
\usepackage[format=plain,justification=justified, singlelinecheck=false,font={stretch=1.125,small,sf},labelfont=bf,labelsep=space]{caption} 
\usepackage{float}
\usepackage{fancyhdr}
\usepackage{fnpos}
\usepackage[english]{babel}

\usepackage{array}
\usepackage{droidsans}
\usepackage{charter}
\usepackage[T1]{fontenc}
\usepackage[usenames,dvipsnames]{xcolor}
\usepackage{setspace}
\usepackage[compact]{titlesec}
\usepackage{hyperref}
\usepackage{physics}
\usepackage{nicefrac}
\usepackage{subcaption}
\usepackage{chemformula}
\usepackage{xspace}

\usepackage{amsmath}
\usepackage{amssymb}
\usepackage{amscd}
\usepackage[np,autolanguage]{numprint}
\usepackage{bm}

  \usepackage{hyperref}
  \usepackage{cleveref}    
  \usepackage{braket}      
  \usepackage{upgreek}     
  \usepackage{chemformula} 
  \usepackage{booktabs}    
  \Crefname{equation}{Eq.}{Eqs.}
  \Crefname{chapter}{Chapter}{Chapters}
  \Crefname{section}{Section}{Sections}
  \Crefname{figure}{Figure}{Figures}
  \Crefname{tabular}{Table}{Tables}
  \Crefname{table}{Table}{Tables}
  \hypersetup{
    colorlinks=true,
    citecolor=blue,
    linkcolor=blue,
    filecolor=magenta,      
    urlcolor=cyan,
  }

\definecolor{cream}{RGB}{222,217,201}

\begin{document}

\pagestyle{fancy}
\thispagestyle{plain}
\fancypagestyle{plain}{
\renewcommand{\headrulewidth}{0pt}
}

\makeFNbottom
\makeatletter
\renewcommand\LARGE{\@setfontsize\LARGE{15pt}{17}}
\renewcommand\Large{\@setfontsize\Large{12pt}{14}}
\renewcommand\large{\@setfontsize\large{10pt}{12}}
\renewcommand\footnotesize{\@setfontsize\footnotesize{7pt}{10}}
\makeatother

\renewcommand{\thefootnote}{\fnsymbol{footnote}}
\renewcommand\footnoterule{\vspace*{1pt}%
\color{cream}\hrule width 3.5in height 0.4pt \color{black}\vspace*{5pt}}
\setcounter{secnumdepth}{5}

\makeatletter
\renewcommand\@biblabel[1]{#1}
\renewcommand\@makefntext[1]%
{\noindent\makebox[0pt][r]{\@thefnmark\,}#1}
\makeatother
\renewcommand{\figurename}{\small{Fig.}~}
\sectionfont{\sffamily\Large}
\subsectionfont{\normalsize}
\subsubsectionfont{\bf}
\setstretch{1.125} 
\setlength{\skip\footins}{0.8cm}
\setlength{\footnotesep}{0.25cm}
\setlength{\jot}{10pt}
\titlespacing*{\section}{0pt}{4pt}{4pt}
\titlespacing*{\subsection}{0pt}{15pt}{1pt}

\fancyfoot{}
\fancyfoot[LO,RE]{\vspace{-7.1pt}\includegraphics[height=9pt]{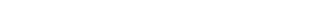}}
\fancyfoot[CO]{\vspace{-7.1pt}\hspace{11.9cm}\includegraphics{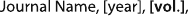}}
\fancyfoot[CE]{\vspace{-7.2pt}\hspace{-13.2cm}\includegraphics{head_foot/RF}}
\fancyfoot[RO]{\footnotesize{\sffamily{1--\pageref{LastPage} ~\textbar  \hspace{2pt}\thepage}}
}
\fancyfoot[LE]{\footnotesize{\sffamily{\thepage~\textbar\hspace{4.65cm} 1--\pageref{LastPage}}
}}
\fancyhead{}
\renewcommand{\headrulewidth}{0pt}
\renewcommand{\footrulewidth}{0pt}
\setlength{\arrayrulewidth}{1pt}
\setlength{\columnsep}{6.5mm}
\setlength\bibsep{1pt}

\makeatletter
\newlength{\figrulesep}
\setlength{\figrulesep}{0.5\textfloatsep}

\newcommand{\topfigrule}{\vspace*{-1pt}%
\noindent{\color{cream}\rule[-\figrulesep]{\columnwidth}{1.5pt}} }

\newcommand{\botfigrule}{\vspace*{-2pt}%
\noindent{\color{cream}\rule[\figrulesep]{\columnwidth}{1.5pt}} }

\newcommand{\dblfigrule}{\vspace*{-1pt}%
\noindent{\color{cream}\rule[-\figrulesep]{\textwidth}{1.5pt}} }

\makeatother

\twocolumn[
  \begin{@twocolumnfalse}
{\includegraphics[height=30pt]{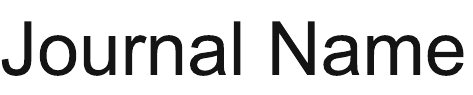}\hfill\raisebox{0pt}[0pt][0pt]{\includegraphics
[height=55pt]{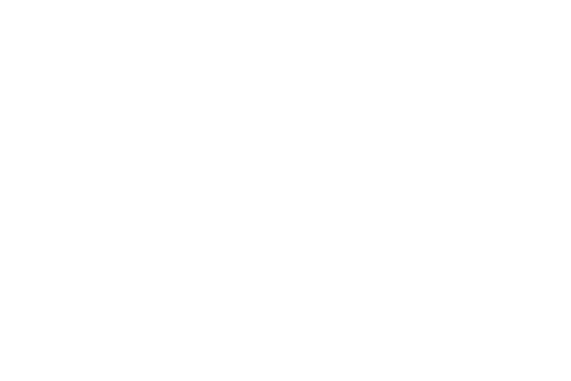}}\\[1ex]
\includegraphics[width=18.5cm]{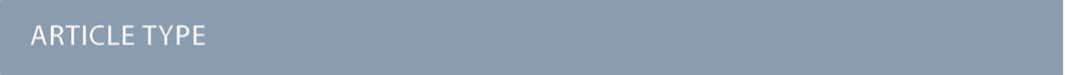}}\par
\vspace{1em}
\sffamily
\begin{tabular}{m{4.5cm} p{13.5cm} }

\includegraphics{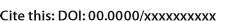} & \noindent\LARGE{\textbf{WTMAD-4: A Fair Weighting Scheme for GMTKN55$^\dag$}} \\
\vspace{0.3cm} & \vspace{0.3cm} \\


 & \noindent\large{Kyle R. Bryenton\textit{$^{a,b}$} and Erin R.\ Johnson\textit{$^{a,b,c\ast}$}} \\


    


\includegraphics{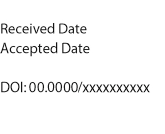} & \noindent\normalsize{%
The GMTKN55 data set is a collection of standard benchmarks used in molecular quantum chemistry that spans small- and large-molecule thermochemistry, reaction barriers, and non-covalent interactions. Herein, we identify a flaw in the weighted mean absolute deviation (WTMAD) definitions commonly used to quantify performance of various electronic-structure methods for the GMTKN55 set, which under-weight some of its component benchmarks by orders of magnitude. A new \mbox{WTMAD-4} metric is proposed, based on typical errors observed for a set of ten minimally empirical dispersion-corrected density-functional approximations (DFAs), ensuring fair treatment across all benchmarks. The performance of 115 DFAs is then reassessed using \mbox{WTMAD-4} and we highlight a literature example where a DFA parametrised by minimising \mbox{WTMAD-2} underperforms for benchmarks marginalised by that metric.
} \\

\end{tabular}

 \end{@twocolumnfalse} \vspace{0.6cm}

  ]

\renewcommand*\rmdefault{bch}\normalfont\upshape
\rmfamily
\section*{}
\vspace{-1cm}


\footnotetext{\textit{$^{a}$~Department of Physics and Atmospheric Science, Dalhousie University, 6310 Coburg Road, Halifax, Nova Scotia, Canada, B3H 4R2}}

\footnotetext{\textit{$^{b}$~Department of Chemistry, Dalhousie University, 6243 Alumni Crescent, Halifax, Nova Scotia, B3H 4R2, Canada. E-mail: erin.johnson@dal.ca}}

\footnotetext{\textit{$^{c}$~Yusuf Hamied Department of Chemistry, University of Cambridge, Lensfield Road, Cambridge, CB2 1EW, United Kingdom.}}

\footnotetext{\dag~Electronic Supplementary Information (ESI) available, see DOI: 10.1039/cXCP00000x/}




\section{Introduction}


Density-functional theory (DFT) methods are a pillar of modern computational chemistry. A principle goal in DFT development is to create functionals that are broadly accurate, transferable, and maintain a low computational cost. However, to assess their performance, thorough benchmarking of DFT methods against reliable experimental or correlated-wavefunction reference data sets is required.\cite{burns2014appointing,kodrycka2019platinum} Since considering a single benchmark may lead to biased results, compiled databases consisting of numerous, chemically diverse benchmarks are now becoming commonplace.\cite{goerigk2017look,liang2025gold,mardirossian2017thirty} The major challenges in creating such benchmarks are obtaining suitable high-level reference data from theory or experiment, curating them to represent a meaningful subset of chemical space, and devising a fair weighting of the component data sets. With respect to this last point, factors such as the wide range of energy scales, quality of the reference data, relative chemical importance, and numbers of systems comprising each benchmark, all become confounding variables.

The GMTKN55 database\cite{goerigk2017look} is a collection of 55 individual benchmarks spanning the thermochemistry of small and large molecules, reaction barriers, and both intramolecular and intermolecular non-covalent interactions (NCI). It has become a standard benchmark in DFT method development in recent years \cite{mehta2018semi,santra2021exploring, wittmann2023dispersion, becke2024remarkably, teale2022dft} due to the robust quality of the reference data and the diverse span of chemical systems. 
The interested reader is directed to Ref.~\citenum{goerigk2017look} for detailed information regarding the individual benchmarks. Due to the wide range of energy scales of its components, the overall error for the GMTKN55 set is reported as a weighted mean absolute deviation (WTMAD), and several definitions for such a weighted error have been proposed. 

The first proposed metric,\cite{goerigk2017look} \mbox{WTMAD-1}, is not commonly used. In this scheme, each subset is assigned an arbitrary weight, denoted $w_{i}$, where $w_{i}=10$ when  $\overline{|\Delta E|}_{i} < 7.5 \text{ kcal/mol}$, $w_{i}=0.1$ when  $\overline{|\Delta E|}_{i} > 75 \text{ kcal/mol}$, and $w_{i}=1$ otherwise. The \mbox{WTMAD-1} is then calculated as
\begin{equation} \label{eq:wtmad1}
\text{WTMAD-1} = \frac{1}{N_\text{bench}} \sum_{i=1}^{N_\text{bench}} w_{i} \cdot \text{MAD}_{i} \,,
\end{equation}
where the sum runs over all $N_\text{bench}=55$ component benchmarks, and MAD$_i$ is the mean absolute deviation between the computed and reference data for the $i$th benchmark. Alternatively, the most widely used error metric for the GMTKN55 is the \mbox{WTMAD-2}, also introduced in Ref.~\citenum{goerigk2017look}, and defined as  
\begin{equation}
\text{WTMAD-2} = \sum_{i=1}^{N_\text{bench}} \frac{N_{i}}{N_{\text{total}}} \cdot \frac{ {\overline{|\Delta E|}_{\text{mean}}}  }{\overline{|\Delta E|}_{i}} \cdot \text{MAD}_{i} \,.
\end{equation}
Here, $N_i$ is the number of data points in the $i$th benchmark and $N_{\text{total}} = \sum_{i=1}^{N_\text{bench}} N_{i} = 1505$ is the total number of data points. $\overline{|\Delta E|}_{i}$ is the average reference energy for the $i$th benchmark, and $\overline{|\Delta E|}_{\text{mean}}$ is the average of all $\overline{|\Delta E|}_{i}$ values.

Error metrics typically are reported for seven categories: ``Basic + Small'', ``Iso + Large'', ``Barriers'', ``Intermolecular NCI'', ``Intramolecular NCI'', ``All NCI'', and GMTKN55 as a whole. When calculating the per-category values, a fixed canonical value of ${\overline{|\Delta E|}_{\text{mean}}} = 56.84~\text{kcal/mol}$ is used across all categories, while $N_\text{total}$ is rescaled per category.  However, due to improvements in the reference data, the value of ${\overline{|\Delta E|}_{\text{mean}}}$ has changed (perhaps unknowingly) from paper to paper, which has been a source of confusion in the literature.\cite{santra2021exploring} 

Previous studies have shown a disproportionate weighting of certain benchmarks within GMTKN55 when computing the WTMAD values. In particular, data in Ref.~\citenum{santra2021exploring} (ESI, Tables S5 to S16) shows that the same benchmarks (such as BH76) were disproportionally weighted very highly across all functionals considered, while contributions from others (such as IL16) were two or more orders of magnitude smaller. This issue is partially due to the fact that some GMTKN55 subsets contain nearly one hundred reaction energies, while others contain fewer than ten. Because the \mbox{WTMAD-2} weights the individual MAD$_i$ values by the number of component reactions, it introduces a bias towards methods that perform well for the larger subsets. Consequently, Wittmann \textit{et al.\@} recently proposed the \mbox{WTMAD-3} weighting scheme, which is identical to \mbox{WTMAD-2} except that it attenuates the weights for a subset to be no more than 1\% of the total reactions considered.\cite{wittmann2023dispersion} The \mbox{WTMAD-3} is calculated via
\begin{equation}\label{eq:wtmad3}
\text{WTMAD-3} = \sum_{i=1}^{N_\text{bench}} \frac{N_{i}^{\text{damp}}}{N_{\text{total}}} \cdot \frac{ {\overline{|\Delta E|}_{\text{mean}}}  }{\overline{|\Delta E|}_{i}} \cdot \text{MAD}_{i} \,,
\end{equation}
where $N_i^{\text{damp}} = \min(0.01 \times N_{\text{total}} \, , \, N_i)$. Thus, for the full GMTKN55, this limits each subset’s weight to 15 systems. However, as we will see, this reweighting does not fully correct for the disproportionate weighting of some GMTKN55 subsets.

A very recent benchmarking study by Liang and Head-Gordon defined a new error metric, the normalised error ratio (NER).\cite{liang2025gold} The NER of a particular functional for a given benchmark is the ratio of its error to the standard error for that benchmark. The NER values are then averaged over all component benchmarks to obtain a value for the full data set, ensuring each has equal weighting. However, the drawback of this approach is that the standard error is defined as the ``average of the second, third, and fourth lowest errors among all tested hybrid functionals.''\cite{liang2025gold} This means that the NER results must be adjusted for each individual benchmarking study to identify the second through fourth lowest errors. Additionally, it is possible that some of the functionals considered are overfit and yield abnormally low errors for certain benchmarks; thus, the standard errors for each individual benchmark may not be representative of what a typical or expected error would be.

In this work, the \mbox{WTMAD-4} metric is proposed. While identical to \mbox{WTMAD-1} in its construction, the weights are based on the magnitudes of expected errors for a small set of representative, minimally empirical functionals, rather than on the absolute energy scales. As a result, each benchmark contributes meaningfully and appropriately to the overall \mbox{WTMAD-4}. Data from previous DFA benchmarking studies \cite{goerigk2017look, santra2021exploring, wittmann2023dispersion, becke2024remarkably} is reassessed using WTMAD-4. 

\section{Data Set} \label{ss:data}

Data for 115 previously studied DC--DFAs\cite{goerigk2017look, santra2021exploring, wittmann2023dispersion, becke2024remarkably} was obtained by extracting the MAD values reported for each of the 55 benchmarks within GMTKN55. These MAD values were then combined with the appropriate weights to generate WTMAD-$N$ values, tabulated for all 115 functionals (and 78 non-dispersion-corrected functionals from Ref~\citenum{goerigk2017look}) in the ESI. Because the complete benchmark results for these 115 previously studied DC--DFAs are not available, and in many cases the specific reference values used are uncertain, their \mbox{WTMAD-$N$} values should be regarded as approximate, with an estimated uncertainty of ca.~1--2\%. Such a small margin of error is not expected to meaningfully impact the statistics presented in this article.

To ensure all previously studied DC--DFAs were treated as consistently as possible, their \mbox{WTMAD-2} and \mbox{WTMAD-3} values were recomputed by combining the MADs reported in earlier work with the updated ${\overline{|\Delta E|}}_i$ values from the revised GMTKN55 reference data,\cite{wittmann2023dispersion, gmtkn55databaseNEW} corresponding to ${\overline{|\Delta E|}_{\text{mean}}} = 57.82~\text{kcal/mol}$.\cite{gmtkn55databaseNEW} However, for consistency with previous works,\cite{goerigk2017look,santra2019minimally} a fixed value of ${\overline{|\Delta E|}_{\text{mean}}} = 56.84~\text{kcal/mol}$ is used when calculating \mbox{WTMAD-2} and \mbox{WTMAD-3}. To obtain \mbox{WTMAD$_{57.82}$} values, an appropriate multiplicative scale factor may be used. We note that neither \mbox{WTMAD-1}, nor the proposed \mbox{WTMAD-4}, depend on ${\overline{|\Delta E|}_{\text{mean}}}$, and thus do not suffer from this issue.

\section{Results and Discussion}

\subsection{Analysis of WTMAD Contributions}

\begin{table}[t!]
\centering
\caption[Contributions to WTMAD values for GMTKN55 subsets]{
Shown for all the GMTKN55 benchmarks are the numbers of component reactions ($N_i$), mean reference energies ($\overline{|\Delta E|}_i$), mean MADs across all 115 DC-DFAs ($\overline{\text{MAD}}_i$), and recommended WTMAD-4 weights ($w_i$). 
Using the data in columns 2--5, and with respective weights from Eqs.\ 1--5, the percent contributions of each benchmark to the total WTMAD-$N$ were evaluated and are shown in the final four columns. Note that the contributions for each subset are not exactly equal for WTMAD-4 as the weights were determined using a much smaller data set of ten minimally empirical functionals.}
\label{tab:wtmadpercent}
{\scriptsize
\setlength{\tabcolsep}{5pt}
\begin{tabular}{lrrrr|rrrr}\hline 
          &     &        &       &      & \multicolumn{4}{c}{WTMAD-$N$} \\
Benchmark & $N_i$ & $\overline{|\Delta E|}_i$ & $\overline{\text{MAD}}_i$ & $w_i$ & 1 &  2 &  3 &  4  \\ \hline
\multicolumn{9}{c}{Basic + Small}                                   \\ \hline
AL2X6     & 6   & 35.88  & 2.38  & 3.27 & 1.20 & 0.24 & 0.43 & 2.25 \\
ALK8      & 8   & 62.60  & 4.91  & 1.30 & 2.48 & 0.37 & 0.68 & 1.85 \\
ALKBDE10  & 10  & 100.69 & 5.09  & 1.06 & 0.26 & 0.30 & 0.55 & 1.55 \\
BH76RC    & 30  & 21.39  & 2.23  & 2.55 & 1.13 & 1.86 & 1.69 & 1.64 \\
DC13      & 13  & 54.98  & 7.22  & 0.78 & 3.64 & 1.02 & 1.84 & 1.63 \\
DIPCS10   & 10  & 654.26 & 4.64  & 1.65 & 0.23 & 0.04 & 0.08 & 2.21 \\
FH51      & 51  & 31.01  & 2.26  & 2.39 & 1.14 & 2.21 & 1.18 & 1.56 \\
G21EA     & 25  & 33.62  & 2.69  & 2.15 & 1.36 & 1.19 & 1.30 & 1.67 \\
G21IP     & 36  & 257.61 & 3.30  & 1.63 & 0.17 & 0.27 & 0.21 & 1.55 \\
G2RC      & 25  & 51.26  & 4.49  & 1.12 & 2.27 & 1.30 & 1.42 & 1.46 \\
HEAVYSB11 & 11  & 58.02  & 3.22  & 2.73 & 1.63 & 0.36 & 0.66 & 2.54 \\
NBPRC     & 12  & 27.71  & 2.06  & 2.32 & 1.04 & 0.53 & 0.96 & 1.38 \\
PA26      & 26  & 189.05 & 2.74  & 1.98 & 0.14 & 0.22 & 0.24 & 1.57 \\
RC21      & 21  & 35.70  & 3.50  & 1.45 & 1.77 & 1.22 & 1.59 & 1.46 \\
SIE4x4    & 16  & 33.72  & 13.08 & 0.47 & 6.60 & 3.69 & 6.30 & 1.77 \\
TAUT15    & 15  & 3.05   & 1.06  & 5.80 & 5.35 & 3.10 & 5.63 & 1.78 \\
W4‐11     & 140 & 306.91 & 5.69  & 1.07 & 0.29 & 1.54 & 0.30 & 1.76 \\
YBDE18    & 18  & 49.28  & 2.81  & 2.55 & 1.42 & 0.61 & 0.93 & 2.07 \\
\multicolumn{5}{r|}{Total:} & 32.12 & 20.07 & 25.99 & 31.70         \\ \hline
\multicolumn{9}{c}{Iso + Large}                                     \\ \hline
BSR36     & 36  & 16.20  & 2.22  & 2.03 & 1.12 & 2.93 & 2.23 & 1.30 \\
C60ISO    & 9   & 98.25  & 5.89  & 1.19 & 0.30 & 0.32 & 0.58 & 2.03 \\
CDIE20    & 20  & 4.06   & 0.96  & 5.75 & 4.85 & 2.81 & 3.84 & 1.59 \\
DARC      & 14  & 32.47  & 3.59  & 1.44 & 1.81 & 0.92 & 1.67 & 1.50 \\
ISO34     & 34  & 14.57  & 1.26  & 4.49 & 0.64 & 1.75 & 1.40 & 1.63 \\
ISOL24    & 24  & 21.92  & 3.15  & 1.99 & 1.59 & 2.05 & 2.33 & 1.81 \\
MB16‐43   & 43  & 468.39 & 21.30 & 0.33 & 1.08 & 1.16 & 0.74 & 2.02 \\
PArel     & 20  & 4.63   & 1.08  & 5.51 & 5.45 & 2.78 & 3.79 & 1.72 \\
RSE43     & 43  & 7.60   & 1.43  & 5.27 & 0.72 & 4.81 & 3.06 & 2.18 \\
\multicolumn{5}{r|}{Total:} & 17.56 & 19.53 & 19.64 & 15.78         \\ \hline
\multicolumn{9}{c}{Barriers}                                        \\ \hline
BH76      & 76  & 18.61  & 4.07  & 1.53 & 2.05 & 9.89 & 3.55 & 1.79 \\
BHDIV10   & 10  & 45.33  & 3.53  & 1.66 & 1.78 & 0.46 & 0.84 & 1.69 \\
BHPERI    & 26  & 20.87  & 2.71  & 2.29 & 1.37 & 2.01 & 2.11 & 1.79 \\
BHROT27   & 27  & 6.27   & 0.45  &10.95 & 2.27 & 1.15 & 1.17 & 1.42 \\
INV24     & 24  & 31.85  & 1.42  & 4.41 & 0.72 & 0.64 & 0.72 & 1.81 \\
PX13      & 13  & 33.36  & 4.60  & 1.18 & 2.32 & 1.07 & 1.93 & 1.56 \\
WCPT18    & 18  & 34.99  & 3.29  & 1.77 & 1.66 & 1.01 & 1.53 & 1.68 \\
\multicolumn{5}{r|}{Total:} & 12.17 & 16.23 & 11.85 & 11.74         \\ \hline
\multicolumn{9}{c}{Intermolecular NCI}                              \\ \hline
ADIM6     & 6   & 3.36   & 0.37  &36.78 & 1.87 & 0.39 & 0.71 & 3.93 \\
AHB21     & 21  & 22.49  & 0.78  & 6.24 & 0.39 & 0.43 & 0.56 & 1.41 \\
CARBHB12  & 12  & 6.04   & 0.87  & 5.50 & 4.39 & 1.03 & 1.87 & 1.38 \\
CHB6      & 6   & 26.79  & 1.52  & 4.14 & 0.77 & 0.20 & 0.37 & 1.82 \\
HAL59     & 59  & 4.59   & 0.70  & 9.63 & 3.53 & 5.35 & 2.48 & 1.95 \\
HEAVY28   & 28  & 1.24   & 0.44  &13.96 & 2.22 & 5.91 & 5.76 & 1.77 \\
IL16      & 16  & 109.04 & 0.64  &11.49 & 0.03 & 0.06 & 0.10 & 2.12 \\
PNICO23   & 23  & 4.27   & 0.69  & 7.86 & 3.48 & 2.21 & 2.63 & 1.56 \\
RG18      & 18  & 0.58   & 0.25  &29.88 & 1.26 & 4.62 & 7.00 & 2.16 \\
S22       & 22  & 7.30   & 0.39  &13.54 & 1.97 & 0.70 & 0.87 & 1.52 \\
S66       & 66  & 5.47   & 0.32  &18.39 & 1.62 & 2.30 & 0.95 & 1.70 \\
WATER27   & 27  & 81.14  & 3.78  & 1.43 & 0.19 & 0.75 & 0.76 & 1.56 \\
\multicolumn{5}{r|}{Total:} & 21.72 & 23.95 & 24.06 & 22.88         \\ \hline
\multicolumn{9}{c}{Intramolecular NCI}                              \\ \hline
ACONF     & 15  & 1.83   & 0.14  &61.19 & 0.71 & 0.68 & 1.24 & 2.47 \\
Amino20x4 & 80  & 2.44   & 0.27  &22.65 & 1.36 & 5.27 & 1.80 & 1.77 \\
BUT14DIOL & 64  & 2.80   & 0.25  &30.02 & 1.26 & 3.40 & 1.45 & 2.17 \\
ICONF     & 17  & 3.27   & 0.29  &20.01 & 1.46 & 0.90 & 1.44 & 1.68 \\
IDISP     & 6   & 14.22  & 2.49  & 2.50 & 1.26 & 0.62 & 1.13 & 1.80 \\
MCONF     & 51  & 4.97   & 0.40  &21.21 & 2.02 & 2.44 & 1.31 & 2.45 \\
PCONF21   & 18  & 1.62   & 0.71  & 8.66 & 3.58 & 4.69 & 7.12 & 1.77 \\
SCONF     & 17  & 4.60   & 0.40  &19.05 & 2.02 & 0.88 & 1.41 & 2.20 \\
UPU23     & 23  & 5.72   & 0.55  &10.12 & 2.78 & 1.32 & 1.56 & 1.61 \\
\multicolumn{5}{r|}{Total:} & 16.45 & 20.20 & 18.46 & 17.92         \\ \hline
\end{tabular}
}
\end{table}

To determine the contributions of each subset to the overall WTMAD-$N$ values,  we first calculated the mean MAD ($\overline{\text{MAD}}_i$) for each GMTKN55 subset by averaging the MAD values across the set of 115 dispersion-corrected functionals. These mean MADs were then used to compute a single representative WTMAD-$N$ value via Eqns.~\ref{eq:wtmad1}-\ref{eq:wtmad3}. Using these WTMAD-$N$'s and the $\overline{\text{MAD}}_i$'s, the percent contributions of each component benchmark to the overall WTMAD-1, WTMAD-2, and WTMAD-3 were evaluated for the entire GMTKN55. 
The results are collected in Table~\ref{tab:wtmadpercent} and show that none of these treatments fairly weight all subsets.

All three previously-proposed WTMAD schemes result in a number of benchmarks having a disproportionately large contribution to the overall WTMAD, while others have near-zero contribution. The largest disparity was observed for \mbox{WTMAD-2}, where BH76 contributes 9.89\% of the total, while DIPCS10 and IL16 contribute only 0.04\% and 0.06\%, respectively; this is a ca.\ 200-fold difference in scale. Upon review, we determined that calculating weights based on the MAD relative to the reference energy, $\overline{|\Delta E|}_{i}$, was not representative. For example, IL16's average reference energy is 109.04 kcal/mol, which is 170 times larger than its mean MAD across our methods of 0.64 kcal/mol. There also exist cases where the inverse is true and the ratio is small, causing over-contribution from subsets such as BH76, which has an average reference energy of 18.61 kcal/mol and a mean MAD of 4.07 kcal/mol. Although dynamic weights allow for easy expansion and transferability to other benchmarks, we could not find a dynamic weighting scheme that fairly represented all systems within GMTKN55. 


\subsection{Formulation of WTMAD-4}

Our analysis of the disparities in \mbox{WTMAD-2} led to the formulation of \mbox{WTMAD-4}, defined as
\begin{equation}
\text{WTMAD-4} = \frac{1}{N_\text{bench}} \sum_{i=1}^{N_\text{bench}}  w_i^{\text{WTMAD-4}} \cdot \text{MAD}_{i} \,.
\end{equation}
This weighting scheme is identical to \mbox{WTMAD-1} in its construction, but the weights are chosen to be
\begin{equation}\label{eq:weights}
w_i^{\text{WTMAD-4}} = \frac{100}{N_{\text{bench}}} \left( \frac{3.5}{\overline{\text{MAD}}_i}\right)\,,
\end{equation}
such that each benchmark contributes evenly to the overall \mbox{WTMAD-4} on average (i.e.\  $\nicefrac{1}{55}=1.82$\% for each benchmark when the mean MAD values are used). Here, $\overline{\text{MAD}}_i$ is the mean MAD for benchmark $i$ across all functionals in the fit set. A factor of 3.5 was also included to scale the WTMAD-4 for consistency with the typical magnitudes of WTMAD-2.

The definition of WTMAD-4 is similar in spirit to the NER.\cite{liang2025gold} However, instead of comparing to the hybrid functionals that give the second through fourth lowest errors, we compare to the mean MADs for a set of ten representative, minimally empirical hybrid functionals. While the WTMAD-4 weights could be evaluated using the mean MADs from the full set of 115 functionals (and results obtained by so doing are shown in the ESI), we recommend instead obtaining the weights from a small set of well-behaved and numerically stable DFAs that are available in most electronic-structure programs. In this way, the error metric will not require readjusting depending on the set of functionals considered, and new weights can easily be generated if the benchmark is expanded. 

Minimally empirical, dispersion-corrected functionals are selected to avoid any excessively low (or high) errors for a particular benchmark that can result from overfitting. The ten functionals selected to determine the mean MAD values and WTMAD-4 weights via Eqn.~\ref{eq:weights} are: PBE0-D3(BJ), B3LYP-D3(BJ), B3P86-D3(BJ), B3PW91-D3(BJ), PW1PW91-D3(0), MPW1PW91-D3(BJ), BHLYP-D3(BJ), HSE06-D3(BJ), HISS-D3(BJ), and LC-$\omega$hPBE-D3(BJ). This recommended set consists of decades-old functionals with predictable behaviours that are well-documented in the literature. Furthermore, inspection of their largest MADs across the GMTKN55 composite benchmarks shows no extreme outliers, in contrast to several of the other DC-DFAs in the full set of 115 examined in this article.

Overall, if the set of functionals used to evaluate the weights performs well for a particular benchmark, the mean MAD will be low and that set will be heavily weighted. Conversely, if the set performs poorly for a particular benchmark, the mean MAD will be high and that set will be lightly weighted. The results in Table~\ref{tab:wtmadpercent} show no drastic over- or under-weighting of the GMTKN55 benchmarks except for ADIM6 (3.93\%), for which there are some very large outliers in the full 115-DFA set, as will be discussed later. All of the other benchmarks contribute between 1.30-2.54\% of the total WTMAD-4 using the mean MADs for all 115 DC-DFAs considered. 

Table~\ref{tab:wtmadpercent} also shows the resulting percent contributions to the total WTMAD-4 for each  GMTKN55 category. The contributions from WTMAD-4 are fairly similar to WTMAD-1, and notably different from WTMAD-2/3. The increased percent contribution of the ``Basic + Small'' classification between \mbox{WTMAD-2} and \mbox{WTMAD-4} is due to the high proportion of benchmarks within this category. However, we consider this shift in weight towards thermochemistry desirable and representative for general chemistry applications. 



\subsection{Validation using Literature Data}

\begin{figure}[t]
\centering
\includegraphics[width=\columnwidth]{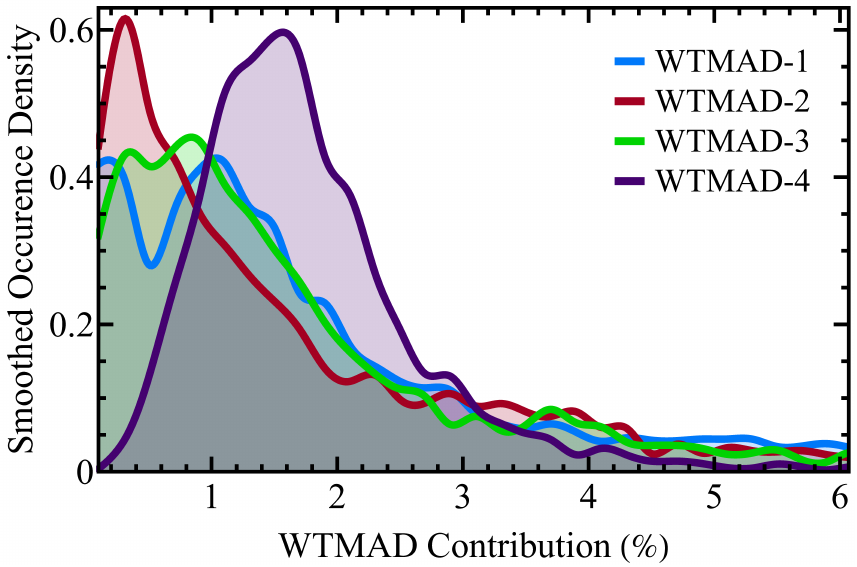}
\caption{Smoothed histograms showing the percent contributions of each GMTKN55 subset to the various WTMAD-$N$ values for 115 DC-DFAs. The WTMAD-4 weights were obtained from Eq.~\ref{eq:weights} using the mean MAD values from ten minimally empirical functionals.}
\label{WTMADHistogram}
\end{figure}

To validate our definition of WTMAD-4, we consider all data available in Refs.~\citenum{goerigk2017look, santra2021exploring, wittmann2023dispersion, becke2024remarkably}, which spans 115 DC-DFAs. Figure~\ref{WTMADHistogram} shows smoothed histograms of the percent contributions to each WTMAD-$N$ obtained for all 55 GMTKN55 data sets with 115 DC-DFAs, where the WTMAD-4 weights were obtained from our recommended subset of 10 well-behaved dispersion-corrected hybrid functionals. The curves show that all of the WTMAD schemes, except WTMAD-4, have many subsets with near-zero contributions and form distributions with slowly decaying tails. WTMAD-1 alone forms a bimodal distribution, likely due to it using only three bins for the weights based on average benchmark energies. Alternatively, WTMAD-4 forms a fast-decaying, skewed distribution that provides a more balanced weighting of all GMTKN55 subsets. A breakdown of the various WTMAD-$N$ contributions by class of DFA is given in the ESI and shows good consistency across GGA, meta-GGA, hybrid, and double-hybrid functionals.

\begin{figure}[t]
\centering
\includegraphics[width=\columnwidth]{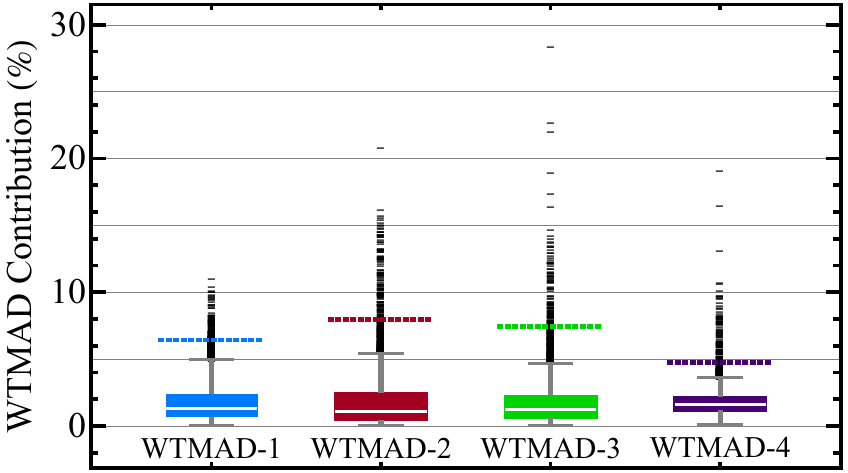}
\caption{Box-and-whisker plots showing the percent contributions of each GMTKN55 subset to the various WTMAD-$N$ values for 115 DC-DFAs. For each WTMAD-$N$, the median is shown with a white line, the coloured boxes span the interquartile range (IQR) from the 25\% to 75\% quantiles. Whiskers extend to 1.5$\times$ the IQR, beyond which outliers are shown as individual points. The 98\% quantile is indicated with a dashed line. The WTMAD-4 weights were obtained from Eq.~\ref{eq:weights} using the mean MAD values from ten minimally empirical functionals.}
\label{f:boxplot}
\end{figure}

The same data shown in Figure~\ref{WTMADHistogram} is recast in a box-and-whisker plot in Figure~\ref{f:boxplot}. Clearly, the WTMAD-4 has the most centralized median and a much tighter inter-quartile range (IQR), indicating a more consolidated spread of contributions towards the total WTMAD. The computed IQRs are 1.70, 1.97, 1.60, 0.97 for WTMAD-1 through WTMAD-4, respectively. Furthermore, the 98th quantile (indicated in Figure~\ref{f:boxplot} using a dashed line) falls at 6.44, 7.98, 7.45, and 4.75\% for WTMAD-1 through WTMAD-4, respectively, again highlighting the latter's more compact distribution.

There are only three large outliers with $>11$\% contributions to the WTMAD-4, which all occur for ADIM6: MN15L-D3(0) (19.1\%), M06-D3(0) (16.5\%), and MN15-D3(BJ) (13.2\%). Moreover, 7 of the 10 outliers with 9--11\% contributions are again for ADIM6. The ADIM6 set corresponds to interaction energies of n-alkane dimers (bound only by London dispersion) and has quite a large WTMAD-4 weight due to the typically small errors ($\overline{\text{MAD}}$ of 0.37 kcal/mol) seen for this benchmark. Analysis of the signed errors shows these three Minnesota functionals systematically overbind the ADIM6 dimers, with MADs of 3.88, 1.82, and 1.32 kcal/mol, for MN15L-D3(0), M06-D3(0), and MN15-D3(BJ), respectively. It was noted in Ref.~\citenum{goerigk2017look} that no DFT-D3(BJ) parameterisation was possible for MN15L due to the overbinding tendency of this highly empirical base functional, and that MN15L-D3(0) was the worst-performing meta-GGA for non-covalent interactions. Additionally, the overall GMTKN55 results did not benefit from the addition of a D3 correction only for M06, MN15L, and MN15, due to their mimicry of dispersion-like binding for van der Waals complexes near equilibrium, although they still neglect long-range London dispersion physics.\cite{goerigk2017look}

Finally, one can consider the scale difference between the minimum-weighted and maximum-weighted GMTKN55 benchmarks for a given DC-DFA. The median scale difference across all 115 functionals is 315.6$\times$ for WTMAD-2, compared to 11.0$\times$ for WTMAD-4. Only five DC-DFAs give a WTMAD-4 scale difference greater than 43.1$\times$, which is the lowest observed scale difference for WTMAD-2 across all methods considered.

\begin{table*}[t!]
\setlength{\tabcolsep}{4pt}
\caption{
The top five (GGA/mGGA) and top ten (Hyb/D.Hyb) functionals for GMTKN55. On the left, sorted by WTMAD-2 values, on the right, sorted by WTMAD-4 values. 
\label{tab:topfunctionals}}
\centering \small
\begin{tabular}{lcc|lcc}\hline
Functional                                                   & WTMAD-2 & WTMAD-4 & Functional                                               & WTMAD-2 & WTMAD-4 \\ 
                                                             & Sorted  &         &                                                          &         & Sorted  \\ \hline
\multicolumn{6}{c}{GGA} \\ \hline
revPBE-D3(BJ)           & 8.29 & 7.73 & revPBE-D3(BJ)           & ~8.29 & 7.73 \\
B97-D3(BJ)              & 8.49 & 8.32 & OLYP-D3(BJ)             & ~8.76 & 8.16 \\
OLYP-D3(BJ)             & 8.76 & 8.16 & B97-D3(BJ)              & ~8.49 & 8.32 \\
BLYP-D3(BJ)             & 9.40 & 8.76 & BLYP-D3(BJ)             & ~9.40 & 8.76 \\
rPW86PBE-D3(BJ)         & 9.67 & 8.94 & MPWPW91-D3(BJ)          & 10.16 & 8.86 \\ \hline
\multicolumn{6}{c}{meta-GGA} \\ \hline
SCAN-D3(BJ)             & 7.96 & 7.95 & revTPSS-D3(BJ)          & 8.44 & 7.73 \\
revTPSS-D3(BJ)          & 8.44 & 7.73 & SCAN-D3(BJ)             & 7.96 & 7.95 \\
M06L-D3(0)              & 8.59 & 8.92 & TPSS-D3(BJ)             & 9.10 & 7.95 \\
TPSS-D3(BJ)             & 9.10 & 7.95 & M06L-D3(0)              & 8.59 & 8.92 \\
M11L-D3(0)              & 9.59 & 9.96 & M11L-D3(0)              & 9.59 & 9.96 \\ \hline
\multicolumn{6}{c}{Hybrid} \\ \hline
$\omega$B97X-V          & 3.93 & 4.40 & $\omega$B97X-V          & 3.93 & 4.40 \\
M052X-D3(0)             & 4.62 & 5.19 & PW6B95-D3(BJ)           & 5.50 & 5.14 \\
$\omega$B97X-D3(0)      & 4.73 & 5.38 & M052X-D3(0)             & 4.62 & 5.19 \\
M062X-D3(0)             & 4.92 & 5.22 & M062X-D3(0)             & 4.92 & 5.22 \\
M08HX-D3(0)             & 5.27 & 5.68 & $\omega$B97X-D3(0)      & 4.73 & 5.38 \\
LC-$\omega$hPBE-D3(BJ)  & 5.44 & 6.43 & MPW1B95-D3(BJ)          & 5.55 & 5.48 \\
PW6B95-D3(BJ)           & 5.50 & 5.14 & M08HX-D3(0)             & 5.27 & 5.68 \\
MPW1B95-D3(BJ)          & 5.55 & 5.48 & MPW1PW91-D3(BJ)         & 6.33 & 5.82 \\
MPWB1K-D3(BJ)           & 5.59 & 6.11 & B1P86-D3(BJ)            & 6.78 & 5.92 \\
BHLYP-D3(BJ)            & 5.65 & 7.02 & PBE0-D3(BJ)             & 6.59 & 5.94 \\ \hline
\multicolumn{6}{c}{Double Hybrid} \\ \hline
DH24                                                  & 1.72 & 2.23 & DH24                                                  & 1.72 & 2.23 \\
revDH23                                               & 1.72 & 2.26 & revDH23                                               & 1.72 & 2.26 \\
XYG8                                                  & 1.85 & 2.66 & SOS-DH24                                              & 1.91 & 2.27 \\
SOS-DH24                                              & 1.91 & 2.27 & SOS-DH23                                              & 1.95 & 2.34 \\
SOS-DH23                                              & 1.95 & 2.34 & $\omega$DOD-PBEP86-D3(BJ) ($\omega=0.10$, $a_X=0.69$) & 2.20 & 2.50 \\
$\omega$DSD-PBEP86-D4     ($\omega=0.16$, $a_X=0.69$) & 2.08 & 2.59 & $\omega$DOD-PBEP86-D3(BJ) ($\omega=0.08$, $a_X=0.72$) & 2.21 & 2.52 \\
$\omega$DSD-PBEP86-D4     ($\omega=0.13$, $a_X=0.72$) & 2.08 & 2.58 & $\omega$DOD-PBEP86-D4     ($\omega=0.10$, $a_X=0.69$) & 2.17 & 2.53 \\
$\omega$DSD-PBEP86-D3(BJ) ($\omega=0.13$, $a_X=0.72$) & 2.10 & 2.57 & $\omega$DOD-PBEP86-D4     ($\omega=0.08$, $a_X=0.72$) & 2.18 & 2.53 \\ 
$\omega$DSD-PBEP86-D3(BJ) ($\omega=0.16$, $a_X=0.69$) & 2.11 & 2.58 & xDOD-PBEP86-D3(BJ)        ($a_X=0.72$)                & 2.23 & 2.53 \\
xDSD-PBEP86-D4            ($a_X=0.75$)                & 2.11 & 2.59 & xDOD-PBEP86-D4            ($a_X=0.72$)                & 2.19 & 2.54 \\ \hline
\end{tabular}
\end{table*}

\subsection{Ranking of DFAs for GMTKN55}


Table~\ref{tab:topfunctionals} shows the top performing of the 115 DC-DFAs considered for the GMTKN55, ranked according to both \mbox{WTMAD-2} and \mbox{WTMAD-4} values for comparison between the two weighting schemes. The results are divided into GGA, meta-GGA, hybrid, and double-hybrid functionals. Included in the ESI are the complete WTMAD-$N$ rankings, as well as full lists of MADs for each benchmark, overall and category-specific partial \mbox{WTMAD-$N$} values, and the percent contributions to \mbox{WTMAD-$N$} for each functional.

From Table~\ref{tab:topfunctionals}, the change in WTMAD scheme results in some order shifting for GGA functionals. While the top four DFAs are consistent, rPW86PBE-D3(BJ) is 5th ranked by \mbox{WTMAD-2} vs.\ 7th by \mbox{WTMAD-4}, while MPWPW91-D3(BJ) is 5th ranked by \mbox{WTMAD-4} vs.\ 9th by \mbox{WTMAD-2}. For the meta-GGAs, there is only slight reordering within the same top five functionals. However, the meta-GGA functionals tend to perform comparatively worse with respect to the GGAs with \mbox{WTMAD-4} than with \mbox{WTMAD-2}. The results show no particular advantage in using the much more complicated meta-GGA functionals, as opposed to the simpler GGAs, for the GMTKN55. 

In contrast, changing the WTMAD definition causes significant re-ordering of the hybrid functionals in Table~\ref{tab:topfunctionals}. PW6B95-D3(BJ) is 7th ranked by \mbox{WTMAD-2} (5.50), but improves to 2nd ranked with \mbox{WTMAD-4} (5.14). Although $\omega$B97X-V remains the top-ranked functional with both metrics, it gives a higher \mbox{WTMAD-4} (4.40) than \mbox{WTMAD-2} (3.93). Examining the MADs of the individual benchmarks for these two functionals, the largest differences arise from the ``Iso.\ + Large'' category. Here, $\omega$B97X-V performs best for benchmarks with large \mbox{WTMAD-2} weights and poorly for those with small weights, while the opposite is true for PW6B95-D3(BJ). In particular, $\omega$B97X-V gives very large MADs for C60ISO and MB16-43 (13.74 and 32.51 kcal/mol), which have quite low \mbox{WTMAD-2} weights. Conversely, PW6B95-D3(BJ) performs excellently for both C60ISO and MB16-43 (MADs of 1.65 and 8.97 kcal/mol), which are more fairly represented by \mbox{WTMAD-4}.

For double hybrid functionals, there is again significant reordering with the change from \mbox{WTMAD-2} to \mbox{WTMAD-4}, which is attributed to the reduced weighting of BH76 and the increased weighting of other benchmarks such as IL16, DIPCS10, ALKBDE10, MB16-43, and WATER27. Notably, WTMAD-2 tends to prefer the spin-component scaled $\omega$DSD double hybrids, while WTMAD-4 favours the $\omega$DOD functionals, in which the same-spin correlation term is omitted, resulting in a lower computational cost.\cite{santra2021exploring} Additionally, Table~\ref{tab:topfunctionals} shows that the XYG8 functional was ranked 3rd by \mbox{WTMAD-2}, but is ranked 17th by WTMAD-4. Compared to other double hybrids, XYG8 underperformed for DIPCS10, ALKBDE10, C60ISO, MB16-43, and WATER27; the increased WTMAD-4 weights for these benchmarks  likely explains the large drop in XYG8's ranking. It is notable that the 8 parameters in XYG8 were fit to  minimise the \mbox{WTMAD-2}.\cite{santra2021surprisingly} In contrast, the empirical parameters in the revDH23 and DH24 functionals were also fit to the GMTKN55 set,\cite{becke2023doubling,becke2024remarkably} but these functionals are the leading performers with both \mbox{WTMAD-2} and \mbox{WTMAD-4}, indicating that they may be more robust.

\section{Summary}

In this work, unintended behaviour of previous WTMAD metrics, used to quantify errors of various electronic-structure methods for the GMTKN55 thermochemical data set, was identified. Specifically, use of energy-based weights in \mbox{WTMAD-2} and \mbox{WTMAD-3} resulted in contributions of some benchmarks being more than 100$\times$ greater than others. The alternative \mbox{WTMAD-4} metric, which uses error-based weights, was proposed herein to ensure that each benchmark within the GMTKN55 set is treated fairly. Ranking the performance of 115 DC-DFAs using \mbox{WTMAD-4} reveals limitations of some functionals through reordering compared to \mbox{WTMAD-2}. Notably, its good performance on the BH76 reaction barriers, at the expense of the DIPCS10, ALKBDE10, C60ISO, MB16-43, and WATER27 benchmarks, resulted in a particularly low \mbox{WTMAD-2} for the XYG8 functional, but a relatively higher \mbox{WTMAD-4}. 

It is tempting to reason that, if particular a method gives a low WTMAD for the GMTKN55, it will perform well for all of its component benchmarks. However, this may not be the case if poor performance for one of more benchmarks is offset by good performance for the other subsets, and it is essential to consider data points that are the largest outliers. Due to unbalanced weighting, fitting to minimise \mbox{WTMAD-2} may result in larger errors for benchmarks with very small contributions, while overfitting to BH76, as seen for XYG8. The more balanced weights in \mbox{WTMAD-4} should reduce the likelihood of, but not eliminate, this possibility. Thus, functional developers should take care to avoid overfitting of functionals to particular thermochemical data and to be cognisant of Goodhart's law that ``when a measure becomes a target, it ceases to be a good measure.''\cite{goodhart1984problems,strathern1997improving} This is particularly important for artificial intelligence (AI) applications of DFT, as well as AI-guided DFA development. It is more sound to consider multiple statistical measures for a data set, rather than focusing on a single number as a target metric.

\section*{Acknowledgements}
KRB and ERJ thank the Natural Sciences and Engineering Research Council (NSERC) of Canada for financial support and the Atlantic Computing Excellence Network (ACENET) for computational resources. ERJ additionally thanks the Royal Society for a Wolfson Visiting Fellowship, while KRB thanks the Killam Trust, the Government of Nova Scotia, and the Mary Margaret Werner Graduate Scholarship Fund.

\section*{Data Availability Statement}
The data that support the findings of this study are available in the supplementary information.

\section*{Conflicts of Interest}

There are no conflicts to report.

\balance

\bibliographystyle{rsc}
\bibliography{Bibliography_2025-09-18.bib}

\providecommand*{\mcitethebibliography}{\thebibliography}
\csname @ifundefined\endcsname{endmcitethebibliography}
{\let\endmcitethebibliography\endthebibliography}{}
\begin{mcitethebibliography}{16}
\providecommand*{\natexlab}[1]{#1}
\providecommand*{\mciteSetBstSublistMode}[1]{}
\providecommand*{\mciteSetBstMaxWidthForm}[2]{}
\providecommand*{\mciteBstWouldAddEndPuncttrue}
  {\def\EndOfBibitem{\unskip.}}
\providecommand*{\mciteBstWouldAddEndPunctfalse}
  {\let\EndOfBibitem\relax}
\providecommand*{\mciteSetBstMidEndSepPunct}[3]{}
\providecommand*{\mciteSetBstSublistLabelBeginEnd}[3]{}
\providecommand*{\EndOfBibitem}{}
\mciteSetBstSublistMode{f}
\mciteSetBstMaxWidthForm{subitem}
{(\emph{\alph{mcitesubitemcount}})}
\mciteSetBstSublistLabelBeginEnd{\mcitemaxwidthsubitemform\space}
{\relax}{\relax}

\bibitem[Burns \emph{et~al.}(2014)Burns, Marshall, and
  Sherrill]{burns2014appointing}
L.~A. Burns, M.~S. Marshall and C.~D. Sherrill, \emph{J. Chem. Phys.}, 2014,
  \textbf{141}, 234111\relax
\mciteBstWouldAddEndPuncttrue
\mciteSetBstMidEndSepPunct{\mcitedefaultmidpunct}
{\mcitedefaultendpunct}{\mcitedefaultseppunct}\relax
\EndOfBibitem
\bibitem[Kodrycka and Patkowski(2019)]{kodrycka2019platinum}
M.~Kodrycka and K.~Patkowski, \emph{J. Chem. Phys.}, 2019, \textbf{151},
  070901\relax
\mciteBstWouldAddEndPuncttrue
\mciteSetBstMidEndSepPunct{\mcitedefaultmidpunct}
{\mcitedefaultendpunct}{\mcitedefaultseppunct}\relax
\EndOfBibitem
\bibitem[Goerigk \emph{et~al.}(2017)Goerigk, Hansen, Bauer, Ehrlich, Najibi,
  and Grimme]{goerigk2017look}
L.~Goerigk, A.~Hansen, C.~Bauer, S.~Ehrlich, A.~Najibi and S.~Grimme,
  \emph{Phys. Chem. Chem. Phys.}, 2017, \textbf{19}, 32184--32215\relax
\mciteBstWouldAddEndPuncttrue
\mciteSetBstMidEndSepPunct{\mcitedefaultmidpunct}
{\mcitedefaultendpunct}{\mcitedefaultseppunct}\relax
\EndOfBibitem
\bibitem[Liang and Head-Gordon(2025)]{liang2025gold}
J.~Liang and M.~Head-Gordon, \emph{arXiv preprint 2508.13468v2}, 2025\relax
\mciteBstWouldAddEndPuncttrue
\mciteSetBstMidEndSepPunct{\mcitedefaultmidpunct}
{\mcitedefaultendpunct}{\mcitedefaultseppunct}\relax
\EndOfBibitem
\bibitem[Mardirossian and Head-Gordon(2017)]{mardirossian2017thirty}
N.~Mardirossian and M.~Head-Gordon, \emph{Mol. Phys.}, 2017, \textbf{115},
  2315–2372\relax
\mciteBstWouldAddEndPuncttrue
\mciteSetBstMidEndSepPunct{\mcitedefaultmidpunct}
{\mcitedefaultendpunct}{\mcitedefaultseppunct}\relax
\EndOfBibitem
\bibitem[Mehta \emph{et~al.}(2018)Mehta, Casanova-P\'{a}ez, and
  Goerigk]{mehta2018semi}
N.~Mehta, M.~Casanova-P\'{a}ez and L.~Goerigk, \emph{Phys. Chem. Chem. Phys.},
  2018, \textbf{20}, 23175--23194\relax
\mciteBstWouldAddEndPuncttrue
\mciteSetBstMidEndSepPunct{\mcitedefaultmidpunct}
{\mcitedefaultendpunct}{\mcitedefaultseppunct}\relax
\EndOfBibitem
\bibitem[Santra \emph{et~al.}(2021)Santra, Cho, and
  Martin]{santra2021exploring}
G.~Santra, M.~Cho and J.~M. Martin, \emph{J. Phys. Chem. A}, 2021,
  \textbf{125}, 4614--4627\relax
\mciteBstWouldAddEndPuncttrue
\mciteSetBstMidEndSepPunct{\mcitedefaultmidpunct}
{\mcitedefaultendpunct}{\mcitedefaultseppunct}\relax
\EndOfBibitem
\bibitem[Wittmann \emph{et~al.}(2023)Wittmann, Neugebauer, Grimme, and
  Bursch]{wittmann2023dispersion}
L.~Wittmann, H.~Neugebauer, S.~Grimme and M.~Bursch, \emph{J. Chem. Phys.},
  2023, \textbf{159}, 224103\relax
\mciteBstWouldAddEndPuncttrue
\mciteSetBstMidEndSepPunct{\mcitedefaultmidpunct}
{\mcitedefaultendpunct}{\mcitedefaultseppunct}\relax
\EndOfBibitem
\bibitem[Becke(2024)]{becke2024remarkably}
A.~D. Becke, \emph{J. Chem. Phys.}, 2024, \textbf{160}, 204118\relax
\mciteBstWouldAddEndPuncttrue
\mciteSetBstMidEndSepPunct{\mcitedefaultmidpunct}
{\mcitedefaultendpunct}{\mcitedefaultseppunct}\relax
\EndOfBibitem
\bibitem[Teale \emph{et~al.}(2022)Teale, Helgaker, Savin, Adamo, Aradi,
  Arbuznikov, Ayers, Baerends, Barone, Calaminici, Canc\`{e}s, Carter,
  Chattaraj, Chermette, Ciofini, Crawford, Proft, Dobson, Draxl, Frauenheim,
  Fromager, Fuentealba, Gagliardi, Galli, Gao, Geerlings, Gidopoulos, Gill,
  Gori-Giorgi, G\"{o}rling, Gould, Grimme, Gritsenko, Jensen, Johnson, Jones,
  Kaupp, K\"{o}ster, Kronik, Krylov, Kvaal, Laestadius, Levy, Lewin, Liu, Loos,
  Maitra, Neese, Perdew, Pernal, Pernot, Piecuch, Rebolini, Reining,
  Romaniello, Ruzsinszky, Salahub, Scheffler, Schwerdtfeger, Staroverov, Sun,
  Tellgren, Tozer, Trickey, Ullrich, Vela, Vignale, Wesolowski, Xu, and
  Yang]{teale2022dft}
A.~M. Teale, T.~Helgaker, A.~Savin, C.~Adamo, B.~Aradi, A.~V. Arbuznikov, P.~W.
  Ayers, E.~J. Baerends, V.~Barone, P.~Calaminici, E.~Canc\`{e}s, E.~A. Carter,
  P.~K. Chattaraj, H.~Chermette, I.~Ciofini, T.~D. Crawford, F.~D. Proft, J.~F.
  Dobson, C.~Draxl, T.~Frauenheim, E.~Fromager, P.~Fuentealba, L.~Gagliardi,
  G.~Galli, J.~Gao, P.~Geerlings, N.~Gidopoulos, P.~M.~W. Gill, P.~Gori-Giorgi,
  A.~G\"{o}rling, T.~Gould, S.~Grimme, O.~Gritsenko, H.~J.~A. Jensen, E.~R.
  Johnson, R.~O. Jones, M.~Kaupp, A.~M. K\"{o}ster, L.~Kronik, A.~I. Krylov,
  S.~Kvaal, A.~Laestadius, M.~Levy, M.~Lewin, S.~Liu, P.-F. Loos, N.~T. Maitra,
  F.~Neese, J.~P. Perdew, K.~Pernal, P.~Pernot, P.~Piecuch, E.~Rebolini,
  L.~Reining, P.~Romaniello, A.~Ruzsinszky, D.~R. Salahub, M.~Scheffler,
  P.~Schwerdtfeger, V.~N. Staroverov, J.~Sun, E.~Tellgren, D.~J. Tozer, S.~B.
  Trickey, C.~A. Ullrich, A.~Vela, G.~Vignale, T.~A. Wesolowski, X.~Xu and
  W.~Yang, \emph{Phys. Chem. Chem. Phys.}, 2022, \textbf{24},
  28700--28781\relax
\mciteBstWouldAddEndPuncttrue
\mciteSetBstMidEndSepPunct{\mcitedefaultmidpunct}
{\mcitedefaultendpunct}{\mcitedefaultseppunct}\relax
\EndOfBibitem
\bibitem[M\"{u}ller and Ehlert(2021)]{gmtkn55databaseNEW}
M.~M\"{u}ller and S.~Ehlert, \emph{gmtkn55}, 2021,
  \url{https://github.com/grimme-lab/GMTKN55},
  https://github.com/grimme-lab/GMTKN55\relax
\mciteBstWouldAddEndPuncttrue
\mciteSetBstMidEndSepPunct{\mcitedefaultmidpunct}
{\mcitedefaultendpunct}{\mcitedefaultseppunct}\relax
\EndOfBibitem
\bibitem[Santra \emph{et~al.}(2019)Santra, Sylvetsky, and
  Martin]{santra2019minimally}
G.~Santra, N.~Sylvetsky and J.~M. Martin, \emph{J. Phys. Chem. A}, 2019,
  \textbf{123}, 5129--5143\relax
\mciteBstWouldAddEndPuncttrue
\mciteSetBstMidEndSepPunct{\mcitedefaultmidpunct}
{\mcitedefaultendpunct}{\mcitedefaultseppunct}\relax
\EndOfBibitem
\bibitem[Santra \emph{et~al.}(2021)Santra, Semidalas, and
  Martin]{santra2021surprisingly}
G.~Santra, E.~Semidalas and J.~M.~L. Martin, \emph{J. Phys. Chem. Lett.}, 2021,
  \textbf{12}, 9368–9376\relax
\mciteBstWouldAddEndPuncttrue
\mciteSetBstMidEndSepPunct{\mcitedefaultmidpunct}
{\mcitedefaultendpunct}{\mcitedefaultseppunct}\relax
\EndOfBibitem
\bibitem[Becke(2023)]{becke2023doubling}
A.~D. Becke, \emph{J. Chem. Phys.}, 2023, \textbf{159}, 241101\relax
\mciteBstWouldAddEndPuncttrue
\mciteSetBstMidEndSepPunct{\mcitedefaultmidpunct}
{\mcitedefaultendpunct}{\mcitedefaultseppunct}\relax
\EndOfBibitem
\bibitem[Goodhart(1984)]{goodhart1984problems}
C.~A. Goodhart, in \emph{Monetary theory and practice: The UK experience},
  Springer, 1984, pp. 91--121\relax
\mciteBstWouldAddEndPuncttrue
\mciteSetBstMidEndSepPunct{\mcitedefaultmidpunct}
{\mcitedefaultendpunct}{\mcitedefaultseppunct}\relax
\EndOfBibitem
\bibitem[Strathern(1997)]{strathern1997improving}
M.~Strathern, \emph{Eur. Rev.}, 1997, \textbf{5}, 305--321\relax
\mciteBstWouldAddEndPuncttrue
\mciteSetBstMidEndSepPunct{\mcitedefaultmidpunct}
{\mcitedefaultendpunct}{\mcitedefaultseppunct}\relax
\EndOfBibitem
\end{mcitethebibliography}

\end{document}